\documentclass[aip,reprint]{revtex4-1}

\usepackage{graphicx}
\usepackage{dcolumn}
\usepackage{bm}
\usepackage{amsmath}
\usepackage{amssymb}
\usepackage[utf8]{inputenc}
\usepackage[T1]{fontenc}
\usepackage{mathptmx}
\usepackage{etoolbox}
\usepackage{color}

\makeatletter
\def\@email#1#2{%
 \endgroup
 \patchcmd{\titleblock@produce}
  {\frontmatter@RRAPformat}
  {\frontmatter@RRAPformat{\produce@RRAP{*#1\href{mailto:#2}{#2}}}\frontmatter@RRAPformat}
  {}{}
}%
\makeatother

\graphicspath{{./}}

\def\Fig{Fig.~}
\def\Figs{Figs.~}
\def\Ref{Ref.~}

\def\Sec{Sec.~}

\def\be{\begin{equation}}
\def\ee{\end{equation}}

\begin{document}
%%%%%%%%%%%%%%%%%%%%%%%%%%%%%%%%%%%%%%%%%%%%%%%%%%%%%%%%%%%%%%%%%%%%%%%%%%%%

% \preprint{AIP/123-QED}

\title[Standalone optical frequency-offset locking electronics]{Standalone optical frequency-offset locking electronics for atomic physics}

\author{K. Shalaby}
\author{T. Hunt}
\author{S. Moir}
\author{P. Trottier}
\author{T. Reuschel}
\author{B. Barrett}
\email[Corresponding author email: ]{brynle.barrett@unb.ca}
\homepage{https://www.quantumsensing.ca}
\affiliation{Dept. of Physics, University of New Brunswick,\\
8 Bailey Dr., Fredericton NB, E3B 5A3, Canada}

\date{\today}

\begin{abstract}
We present a standalone frequency-offset locking system for controlling narrow-linewidth lasers using off-the-shelf electronic components. We lock two frequency-doubled 1560 nm lasers to a stable primary laser operating at 780 nm via their optical beat note. This radio-frequency beat note is fed through a broadband variable divider, a frequency-to-voltage converter, and a proportional-integrator controller to lock each follower laser to a tunable offset frequency relative to the primary. This architecture provides a large capture range ($> 1$ GHz), fast response times ($< 1$ ms), and high linearity. We achieve a frequency resolution of 1.9 kHz and a short-term fractional frequency instability $10^{-11}/\sqrt{\tau \rm (s)}$ at 780 nm without the need for a dedicated, precise clock reference. We perform high-resolution spectroscopy of cold $^{87}$Rb atoms to demonstrate the tunability and precision of our locking system. We designed the system to be modular and extensible, making it applicable to a wide variety of atomic physics experiments, including laser cooling, spectroscopy, and quantum sensing with atoms, ions, and molecules.
\end{abstract}


\maketitle

%%%%%%%%%%%%%%%%%%%%%%%%%%%%%%%%%%%%%%%%%%%%%%%%%%%%%%%%%%%%%%%%%%%%%%%%%%%%
%===========================================================================
\section{Introduction}

Atomic sensors leverage the quantum properties of atoms to measure a broad range of physical quantities with unprecedented sensitivity and accuracy \cite{Kitching2011, Geiger2020, Brown2025}. Many of these features can be traced back to the exceptional coherence properties of lasers, which are a key technological component in these quantum sensors. In most instruments, dynamic control of the optical frequency and/or phase is required to effectively manipulate quantum systems. To reach the desired level of fractional frequency instability (FFI), which is typically $< 10^{-8}$, available techniques involve transferring the coherence of a ``primary'' oscillator to a ``follower'' oscillator using a feedback loop. When the follower laser is stabilized to a frequency differing from that of the primary, the feedback loop is referred to as a frequency-offset lock.

Several frequency-offset locking methods have been demonstrated that transfer full phase coherence between the lasers. Reference \citenum{Keitch2013} used injection locking with an acousto-optic modulator (AOM) between the lasers to achieve frequency offsets up to 3.2 GHz. A high-bandwidth electro-optic modulator (EOM) was used in \Ref \citenum{Peng2014} to achieve up to 40 GHz of frequency offset on either side of the carrier frequency. There, an optical sideband produced by the EOM is locked to a spectroscopic feature and the carrier is controlled by the radio frequency (RF) injected into the EOM. Optical transfer cavities coupled with AOMs or EOMs have been used to realize locks with large tuning ranges for molecular spectroscopy \cite{Biesheuvel2013} and narrow-line laser cooling \cite{Nevsky2013}. Finally, by beating two lasers on a fast photodetector, an optical phase-locked loop (OPLL) can be realized with offset frequencies up to the detector bandwidth (15 GHz in \Ref \citenum{Santarelli1994}). This method relies on an ultra-stable local oscillator (LO) near the offset frequency to mix down the beat note---producing an error signal proportional to $\sin(\Delta \omega t + \Delta \phi)$. Here, $\Delta \omega$ is the frequency difference between the optical beat note and the LO, and $\Delta \phi$ is the corresponding phase difference. The feedback loop then matches the frequency and phase difference between the lasers to that of the LO. With sufficient locking bandwidth, these systems are capable of narrowing a laser's linewidth. OPLLs employing commercial phase-frequency detectors\cite{Lipka2017} have achieved FFIs at the level of $10^{-13}/\sqrt{\tau({\rm s})}$. More recent work has demonstrated frequency jumps up to 1.1 GHz with response times $<100$ $\mu$s when OPLLs are coupled with feed-forward electronics \cite{Han2025}.

Implementing these methods requires expensive hardware (e.g., AOMs, EOMs, transfer cavities, ultra-stable LOs). Robust phase locks also require a large feedback bandwidth (several MHz) and a wide frequency capture range, which is challenging to realize. Although phase locks provide excellent frequency stability, they are often too demanding for applications that require FFIs between $10^{-11} - 10^{-9}$ on a 1 h timescale. In such cases, simpler methods based on frequency-to-voltage converters (FVCs) have demonstrated better than kHz-level frequency stability \cite{Hughes2008, Li2022} without phase stabilization. FVC-based locks have also been used in conjunction with OPLLs to provide increased capture range and robustness \cite{Seishu2019}.

One such scheme uses the steep slope of an RF filter transfer function to realize a sensitive FVC \cite{Ritt2004, Strauss2007, Puentes2012, Cheng2017}. Here, the optical beat note is split into two channels, one of which transmits through an RF filter. The RF amplitude difference between the two channels provides an error signal with a linear dependence on frequency near the zero crossing due to the transfer function of the filter. This scheme provides a single locking point determined by the RF filter characteristics. Dynamic tunability of the lock point is provided by a LO, typically a direct digital synthesizer (DDS), that is mixed with the optical beat note. Recently, \Ref \citenum{Li2022} used a hybrid LC/RLC filter to achieve a capture range of 190 MHz, and a FFI $< 1.3 \times 10^{-13}$ (50 Hz at 780 nm) for timescales between $1 - 1000$ s.

A more direct method involves the use of a commercial FVC integrated circuit (IC), which produces a dc voltage proportional to the input frequency. The IC uses a charge pump that delivers a fixed amount of charge to an integrator on every cycle of the input signal. The average current from these charge packets is proportional to frequency, and the integrator converts that current into a dc voltage. The first frequency-offset lock based on a FVC IC was demonstrated by \Ref \citenum{Stace1998} and reported a FFI $< 2.6 \times 10^{-13}$ (100 Hz at 780 nm) for timescales between $1 - 20$ s. However, their system was limited to offset frequencies below 8 MHz due to detector bandwidth constraints. Upgrades to both the detector and frequency division circuitry have since expanded this range to 1-2 GHz. Several groups have adopted this approach for laser cooling and trapping \cite{McFerran2018, Templier2021}, onboard atom interferometry \cite{Stern2009, Carraz2009, Menoret2011, Dinkelaker2017}, and measuring laser frequency noise \cite{Antona2022}. Yet there has been surprisingly little work that focuses on the design and performance of this architecture.

In this work, we discuss a refined version of the FVC-based frequency-offset lock (henceforth "lock") using low noise, high bandwidth electronics with a modular circuit design. We built two equivalent locking systems and validate their performance using different operating parameters. We demonstrate the dynamic control of two lasers over GHz-scale frequency ranges with ms-scale response times. Compared to other architectures, our approach does not require an ultra-stable clock, expensive active optics, or complex dynamic programming of a DDS---making it self-contained and simple to implement. The trade-off is that the optical frequency can exhibit a non-linear dependence on the external control voltage, and the frequency stability can be limited by the control voltage stability and the temperature sensitivity of internal circuit components. To combat these factors, we selected components with low non-linearity and temperature sensitivity, and we focused on the development of ultra-low-noise electronics.

The remainder of this article is organized as follows. Section \ref{sec:LaserSystem} presents the laser system we used for this study. In \Sec \ref{sec:FOL}, we discuss our FVC and locking circuit design, followed by a characterization of its behaviour and performance in \Sec \ref{sec:LockCharacterization}. Finally, in \Sec \ref{sec:Spectroscopy}, we demonstrate the capability of our lock by performing high-resolution spectroscopy in laser-cooled $^{87}$Rb, and we draw conclusions in \Sec \ref{sec:Conclusion}.

%===========================================================================
\section{Laser System}
\label{sec:LaserSystem}

\begin{figure}[!tb]
  \centering
  \includegraphics[width=0.48\textwidth]{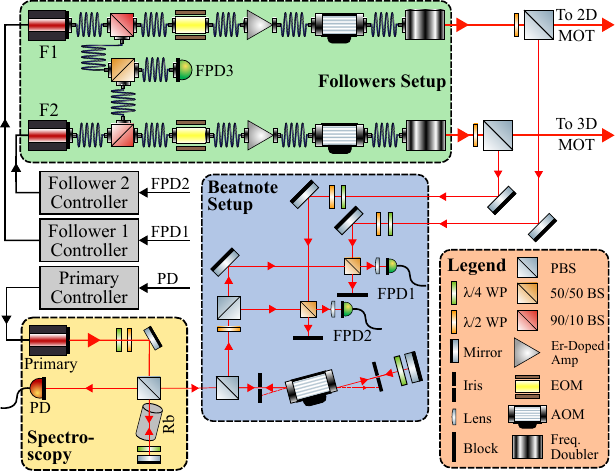}
  \caption{Block diagram showing the different subsystems of our laser setup. F1: follower 1 laser, F2: follower 2 laser, PBS: polarizing beam splitter, $\lambda/4$: quarter-waveplate, $\lambda/2$: half-waveplate, PD: photodiode, FPD: fast photodiode, AOM: acousto-optic modulator, EOM: electro-optic phase modulator, MOT: magneto-optical trap. FPD1 and FPD2 measure the beat notes between the primary laser and follower lasers 1 and 2, respectively. FPD3 monitors the beat note between the two follower lasers.}
  \label{fig:OpticalSetup}
\end{figure}

Our laser system, shown in \Fig \ref{fig:OpticalSetup}, is designed for atom interferometry experiments with laser-cooled $^{87}$Rb atoms. It comprises three lasers sources: a primary laser and two follower lasers. The primary laser is a 40 mW distributed feedback (DFB) diode laser at 780 nm (Toptica Eagleyard EYP-DFB-0780-00040-1500-BFW11-0005) with a typical linewidth of 600 kHz. It is stabilized to the $F=2 \to F'=3$ transition in $^{87}$Rb using saturated absorption spectroscopy \cite{Foot2005} with direct frequency modulation at 100 kHz. Using the frequency sensitivity of the error signal\cite{Shalaby2025}, we estimate a FFI of $3.3 \times 10^{-12}/\sqrt{\tau \rm{(s)}}$ (1.3 kHz at 780 nm). The follower lasers are fiber-coupled diode lasers operating at 1560 nm (Eblana EP1560-5-NLW-B26-200FM), each delivering approximately 8 mW of optical power with a typical linewidth of 200 kHz. The output of each follower laser is divided using a series of fiber splitters. Approximately 10\% of each follower is combined on a 10 GHz photoreceiver (Thorlabs RX10AF, FPD3 in \Fig \ref{fig:OpticalSetup}) for beat note monitoring and linewidth characterization at 1560 nm. The remaining 90\% is routed through an EOM (iXblue MPZ-LN-10-00-P-P-FA-FA) that generates optical sidebands at 6.57 GHz for creating repump light during laser cooling, or 6.83 GHz for driving optical Raman transitions during atom interferometry. This light is then sent to an erbium-doped fiber amplifier (EDFA, Lumibird CEFA-C-BO-HP) that produces up to 5 W of optical power. The amplified light then passes through a fiber-coupled AOM (AA Opto-Electronic MT110-IIR25-Fio-PM5-J1-A-VSF) that shifts the frequency by 110 MHz and provides dynamic power control. Finally, the output of the AOM is sent through a periodically-poled lithium niobate (PPLN) crystal waveguide (Covesion WGCF-1560-40) that converts the wavelength to 780 nm via second-harmonic generation. Each follower setup provides up to 1 W at 780 nm to prepare our cold atom source. Follower 1 is primarily used to generate a two-dimensional magneto-optical trap (2D MOT), while follower 2 provides light for the 3D MOT, sub-Doppler cooling, state preparation, atom interferometry, and fluorescence detection stages of our experiments.

The follower lasers are each locked to the primary with a tunable offset frequency using our FVC-based electronics. To facilitate these locks, the primary is first blue-shifted by $\sim 200$ MHz using a double-pass AOM (AA Opto-Electronic MT80-A1-IR). This avoids a beat frequency of zero when the follower lasers are near the $F=2 \to 3'$ resonance. Light from each follower laser is then spatially overlapped with the frequency-shifted primary on a 50/50 beamsplitter and aligned on a custom-designed fast photodiode (FPD) to generate an optical beat note up to 3 GHz\footnote{Our fast photodiode circuit comprises a Hamamatsu S9055-01 high-speed photodiode and a Mini-Circuits GALI-74+ RF amplifier integrated on a PCB that achieves a noise floor of –120 dBm and a signal-to-noise ratio $> 50$ dB for frequencies up to 3 GHz.}. These beat signals are processed by feedback controllers that regulate the frequency of each follower, ensuring both short-term agility and long-term stability with the primary.

\begin{figure}[!tb]
  \centering
  \includegraphics[width=0.48\textwidth]{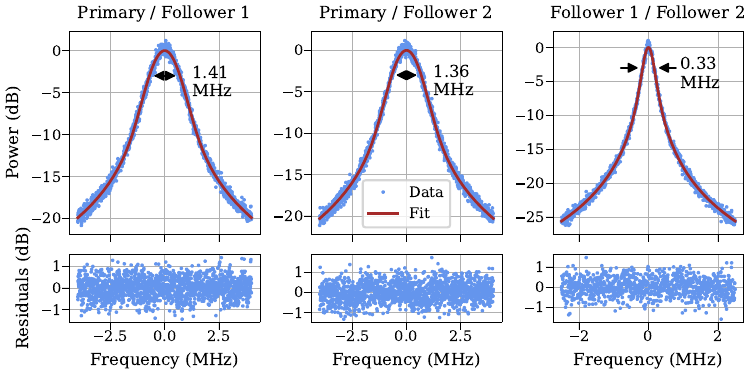}
  \caption{Top: beat note spectra measured between each pair of lasers. Data are fit to a Voigt distribution, which yielded full widths at half maximum of $1.412(9)$ MHz, $1.361(9)$ MHz, and $0.332(7)$ MHz, respectively. Bottom: fit residuals for each beat note spectrum.}
  \label{fig:BeatNoteSpectra}
\end{figure}

Figure \ref{fig:BeatNoteSpectra} shows the measured power spectra of the beat note signals between each pair of lasers. These data were recorded with the primary laser free-running to avoid additional broadening produced by the frequency modulation used to lock the primary. Each beat note spectrum is fit to a Voigt profile to extract its full-width at half-maximum (FWHM). We used a simple model\cite{Shalaby2025} to extract the linewidths of each laser ($\Gamma$) from combinations of the FWHM from the three beat note spectra. We found $\Gamma = 1236.4(9.2)$ kHz for the primary laser, 238(16) kHz for follower 1, and 148(32) kHz for follower 2. These results are consistent with the manufacturer's specifications for the follower lasers. However, the measured linewidth of the primary laser is approximately $2\times$ the typical specification. We attribute this to the elevated current sensitivity of DFB lasers at 780 nm and the current noise generated by our custom laser controllers.

Evidence of additional technical noise in the primary laser is present in the beat note spectra shown in \Fig \ref{fig:BeatNoteSpectra}. Technical noise, such as current noise in diode laser controllers, typically appears as inhomogenous broadening of the laser lineshape---corresponding to a larger Gaussian component of the Voigt profile\cite{Stephan2005, DiDomenico2010}. The remaining Lorentzian component of the Voigt profile provides the laser's linewidth due to homogeous noise sources such as amplified spontaneous emission. Fits to the primary/follower beat note spectra reveal a larger Gaussian component to the linewidth ($\sim 77\%$) compared to the follower 1/follower 2 spectrum ($\sim 50\%$ Gaussian). These linewidth figures are sufficient for this study and can be improved in the future by employing laser controllers with lower current noise density.

%===========================================================================
\section{Frequency-Offset Locking Electronics}
\label{sec:FOL}

\begin{figure}[!tb]
  \centering
  \includegraphics[width=0.48\textwidth]{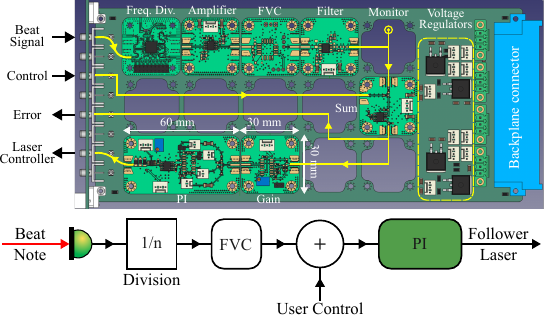}
  \caption{PCB layout (top) and the corresponding block diagram (bottom) of the locking system. PCB tiles for each stage are mounted to a support board that provides power, inputs for the beat note signal and user control voltage, and outputs for the error signal, laser feedback, and FVC monitor.}
  \label{fig:LockCircuit}
\end{figure}

The follower lasers are stabilized to frequencies red-detuned from the primary laser using two separate locking systems, each based on a commercial FVC IC. The electronic design of the lock is illustrated in \Fig \ref{fig:LockCircuit}. It consists of four basic stages: (1) frequency division, which scales down the beat note from GHz to sub-MHz frequencies; (2) frequency-to-voltage conversion, which produces a dc voltage proportional to the divided frequency; (3) signal summation, which combines the FVC output with a user-controlled voltage to generate an error signal; and (4) an analog proportional-integral (PI) controller, the output of which is fed back to the laser. Each of these stages is realized on a separate printed circuit board (PCB) ``tile'', as shown in Figure \ref{fig:LockCircuit}. A support board with a backplane connector and integrated adjustable voltage regulators (onsemi LM317 and LM337) can accommodate several standard tile sizes of $30 \times 30$ mm or $60 \times 30$ mm. This modular architecture greatly simplifies debugging and upgrades, since functional blocks can be tested in isolation, and individual tiles can be replaced or reconfigured with minimal disruption. The support board also includes several capacitive decoupling filters to reduce regulator voltage noise.

\begin{figure}[!tb]
  \centering
  \includegraphics[width=0.48\textwidth]{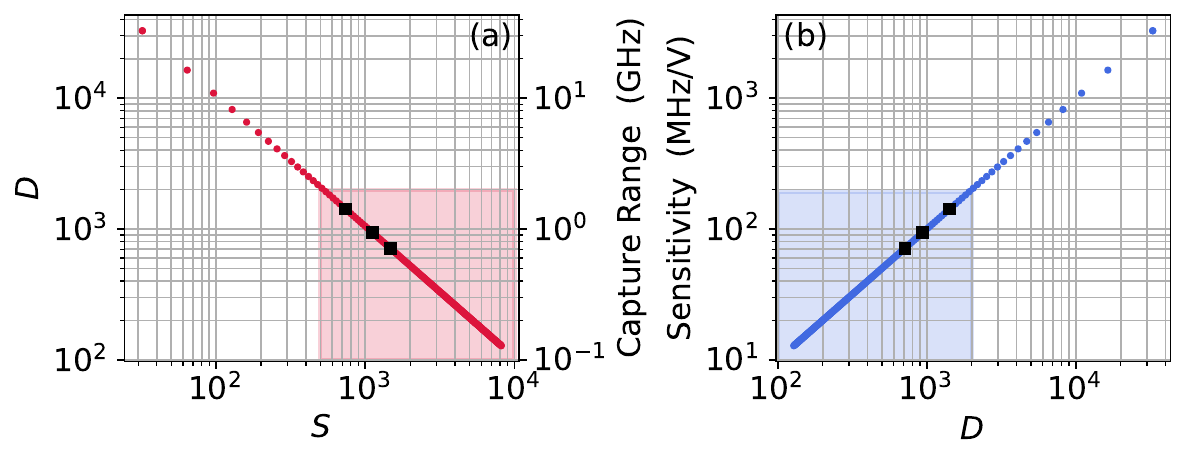}
  \caption{(a) Prescaler division ratios $D$ for the seed values $S$ within our constrained 8-bit range. The approximate capture range of the lock is shown on the right-hand vertical axis. (b) Open-loop frequency sensitivity of the locking system as a function of $D$. In both plots, the shaded regions indicate the range accessible by our laser controllers ($\sim 2$ GHz capture range determined by a $\pm 1 ~\rm{V}$ current modulation range and 1.67 mA/V current sensitivity). Solid squares indicate the division ratios used in this work: $D = 712$, 936, and 1424.}
  \label{fig:DivisionRatios}
\end{figure}

The frequency division stage uses a broadband prescaler (Microsemi MX1DS10P) with a programmable division ratio. This component accepts input frequencies from 50 MHz to 15 GHz and provides a tunable division ratio, $D = 2^{20}/S$, where $S$ is a 20-bit seed value selected via binary switches (maximum value: $S = 2^{19}$). For an input sinusoidal signal at frequency $f_{\rm in}$, the divider produces a $\pm 1$ V square wave at frequency $f_{\rm in}/D$. This output is subsequently buffered by a high-bandwidth operational amplifier (Analog Devices ADA4637, gain-bandwidth product 80 MHz) to isolate the signal from downstream circuitry. We constrain the seed value to an 8-bit subset between $S \simeq 2^5 \to 2^{13}$, which allows us to set the division ratio between $D \simeq 2^7 \to 2^{15}$, as shown in \Fig \ref{fig:DivisionRatios}(a).

The FVC IC (Analog Devices AD650JPZ) was chosen for its autonomy, wide bandwidth, and high degree of linearity. It does not require a clock signal, it supports input frequencies up to 1 MHz (full scale range), and features a typical nonlinearity of 0.1\% (defined as the fractional frequency deviation from linear over the full scale range). A low nonlinearity is important for applications requiring accurate knowledge of the optical frequency (e.g., high-resolution spectroscopy, metrology with atomic clocks or atom interferometers). The FVC stage generates a dc voltage $V_{\rm FVC} = K f_{\rm in}$ between 0 and 10 V, where the input frequency $f_{\rm in} = f_{\rm beat}/D$ is the divided beat frequency, and the proportionality factor $K \simeq 10$ V/MHz is determined by two external capacitors and one resistor. The open-loop sensitivity of the lock is then $\partial f/\partial V = D/K$, which is shown in \Fig \ref{fig:DivisionRatios}(b) as a function of $D$. The frequency capture range of the lock is determined by the maximum output voltage of the FVC ($V_{\rm max} = 10$ V) and the lock sensitivity: $f_{\rm cap} \simeq (\partial f/\partial V) V_{\rm max}$, which is between 0.7 and 1.4 GHz for our operating parameters. We show the capture range as a function of $D$ in \Fig \ref{fig:DivisionRatios}(a). The dynamic range of the lock is determined by the maximum output voltage of the FVC and the resolution of the control voltage. Our control voltage is derived from a 16-bit digital-to-analog converter with a resolution of $\Delta V = 0.31$ mV, yielding a dynamic range up to $10 \log_{10}(V_{\rm max}/\Delta V) = 45$ dB. Practical constraints, such as the current modulation range of a laser controller, can reduce the dynamic range.

We operate the FVC over its full-scale range ($0 - 1$ MHz) by following the manufacturer's recommendations. Similarly, we used an integrator capacitor of $C_{\rm int} = 1$ nF to achieve high bandwidth performance. A known limitation of the FVC is leakage of the input frequency into the output voltage, which appears as a ripple superimposed on the dc signal. To mitigate this effect, we introduced a second-order active low-pass filter with an 8 kHz cut-off and unity gain. Implemented in a Sallen–Key configuration, this filter provides a roll-off of –40 dB/decade that effectively suppresses ripples and dramatically improves the FVC output stability. With this filter in place, voltage ripples in the FVC output are reduced below a root-mean squared (rms) of 0.54 mV. Although this low-pass filter ultimately limits the bandwidth of the lock, it is essential for achieving good stability. Our choice of 8 kHz cut-off represents a balance between lock stability, bandwidth, and dynamic range.

The summation stage sums the filtered FVC output with a $-5~\rm{V}$ reference (to center the FVC output on 0~V) and a user-controlled voltage equal to the negative of the set point: $-V_{\rm set}$. The resulting error signal ($\epsilon = V_{\rm FVC} - V_{\rm set} - 5$~V) is sent to a ``global gain'' stage and then to the PI controller.

The PI controller is implemented with a single quad operational amplifier (Texas Instruments LM324DR) with a 1.2 MHz gain-bandwidth product. The controller includes one proportional stage and two parallel integrator stages with distinct time constants, which we refer to as the fast and slow integrators, respectively. This dual-integrator design allows the servo to simultaneously handle fast disturbances and slow drifts by distributing integration across time scales. This is common practice in high-performance laser servos that are sensitive to wideband noise (e.g., acoustic, electronic) and very slow drift (e.g., thermal expansion). We use integrator time constants of 0.62 ms (fast) and 67 s (slow). Both integrators operate without gain-limiting feedback resistors; their gain is set by the global gain stage prior to the PI circuit. The proportional stage has independent gain control via a feedback resistor. The controller output passes through an analog switch, providing fast and low-noise toggling of the output signal. We measure an output noise floor of 0.54 mV (rms) in open-loop operation and $\sim 1.4$ mV (rms) in closed-loop. Finally, a series of diodes limit the output voltage to $\pm 1$ V to prevent over-driving downstream electronics (e.g., laser current controllers). The PI gains are tuned using a standard impulse-response method in which a square-wave modulation is applied to the control voltage to induce abrupt changes in the lock set point. The global and proportional gains are manually adjusted to achieve minimal settling time. We measure a $-3$ dB lock bandwidth of $f_{\rm bw} = 7.40(1)$ kHz with this method (see \Fig \ref{fig:Agility} below).

Our PCBs are built on industry standard FR-4 substrates following established RF design principles to ensure robust performance and ultra-low noise. The power delivery network contains localized decoupling and filtering for each tile. Surface-mount components are employed throughout to reduce parasitic capacitance and inductance effects. Noise resulting from signal cross-talk and reflections is mitigated using localized reference potentials, grounding vias, and impedance-matched transmission lines. Together, these measures yield a noise floor below $-120$ dBm and enable stable operation up to at least 100 MHz without parasitic resonances. The FPD circuit and frequency divider tile support signals up to 3 GHz and 4 GHz, respectively. Larger bandwidths can be supported by employing precise impedance control and using a substrate with a lower dielectric constant.

%===========================================================================
\section{Lock Characterization}
\label{sec:LockCharacterization}

\begin{figure}[!tb]
  \centering
  \includegraphics[width=0.48\textwidth]{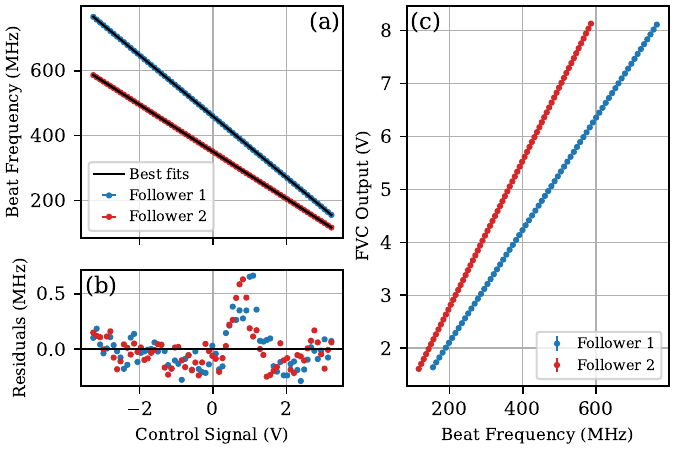}
  \caption{Closed-loop calibration of locking systems for followers 1 and 2. (a) Measurements of the optical beat frequency as a function of the control voltage. Linear fits to these data yield slopes of $-92.572(11)$ MHz/V for follower 1 ($D = 936$) and $-71.238(10)$ MHz/V for follower 2 ($D = 712$). (b) Fit residuals. The excursion near 1~V is due to the FVC nonlinearity, which is $< 0.1\%$ for both systems. (c) FVC output voltage as a function of beat frequency. Linear fits yield slopes of 0.010590(13) V/MHz for follower 1, 0.013864(21) V/MHz for follower 2.}
  \label{fig:LockCalibration}
\end{figure}

We calibrated the closed-loop behaviour of the locking systems by scanning the control voltage and simultaneously measuring the output of the FVC and the optical beat note using a spectrum analyzer. Figure \ref{fig:LockCalibration} shows the central beat frequency (extracted from fits to individual beat note spectra) as a function of the control voltage. The beat frequencies closely follow a linear model as a result of the FVC. We find a maximum nonlinearity of $0.7/944 = 0.074\%$ for the follower 1 lock, and $0.6/721 = 0.083\%$ for follower 2, which are consistent with the FVC specification.

\begin{figure}[!tb]
  \centering
  \includegraphics[width=0.48\textwidth]{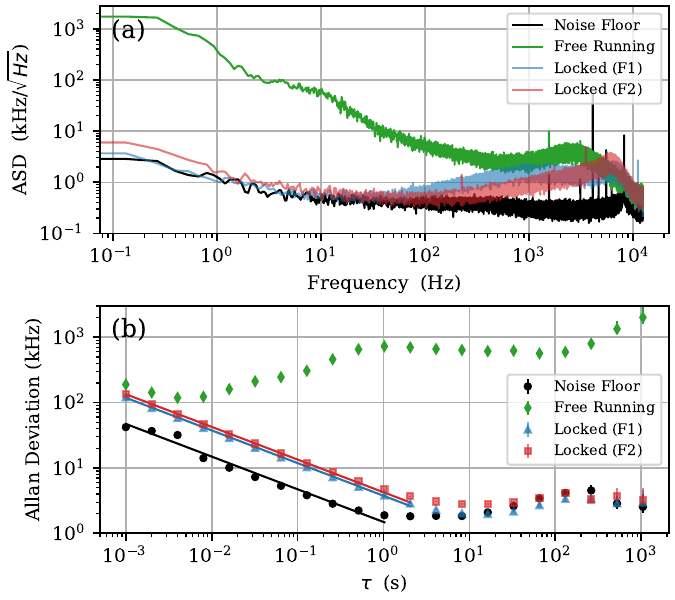}
  \caption{Performance of the locking systems. (a) Amplitude spectral density (ASD) of short-term frequency fluctuations (1 min. acquisition, 25 kS/s). Noise floor of the FVC (black) obtained in open-loop by injecting a stable 550 MHz RF signal. Free-running laser spectrum (green) obtained in open-loop from a $\sim 500$ MHz optical beat note. Laser frequency noise obtained in closed-loop for followers 1 and 2 (F1 in blue and F2 in red). Here, the division ratios were $D = 936$ and $D = 1424$ for followers 1 and 2, respectively, with corresponding open-loop sensitivities $\partial f/\partial V \simeq 93.6$ MHz/V and 142.4 MHz/V. (b) Total Allan deviation of long-term frequency fluctuations (1 hour acquisition, 1 kS/s) for the same cases shown in (a).}
  \label{fig:LockPerformance}
\end{figure}

We also characterized the spectral response and the temporal stability of the locking systems. Figure \ref{fig:LockPerformance}(a) shows the amplitude spectral density (ASD) of frequency fluctuations of each laser in both open- and closed-loop operation. These data were measured directly from the FVC output using a 16-bit data acquisition card (MCC USB-231) and a custom single-ended to differential converter to mitigate ground loops and voltage noise. The noise floor of the lock was obtained in open-loop by inserting a stable 550 MHz signal from an RF generator. It exhibits a quasi-white noise response from 10 Hz to 10 kHz, while lower frequencies follow a $\sim f^{-1/2}$ trend characteristic of flicker frequency noise. The large frequency spurs near 4.1 kHz and 8.2 kHz are produced by the RF generator and are not present in our optical signals.

When free-running, the ASD of both follower lasers exhibits a $\sim f^{-1}$ response for frequencies below 100 Hz characteristic of random-walk frequency noise. This is due to residual thermal drift of the laser cavities and the current noise of our controllers. When locked, both locking systems show excellent noise suppression over this frequency range---reaching a factor 300 at 1 Hz. The follower 1 lock is consistent with the noise floor below 50 Hz, but diverges for larger frequencies---plateauing above 1 kHz due to the finite bandwidth of the PI circuit. Follower 2 remains above the noise floor at low frequencies due to the reduced open-loop sensitivity at larger division factors. Above 100 Hz, follower 2 also diverges due to bandwidth limitations. All data exhibit a cut-off near 8 kHz as a consequence of the FVC output filter.

Figure \ref{fig:LockPerformance}(b) shows the total Allan deviation of optical beat note fluctuations over a 1 h period. Both locking systems exhibit white noise behaviour for integration times $\tau \lesssim 2$ s. Power-law fits to these data yield short-term frequency instabilities of 3.76(2) kHz$/\sqrt{\tau \rm (s)}$ and 4.26(1) kHz$/\sqrt{\tau \rm (s)}$ for followers 1 and 2, respectively---corresponding to FFIs of approximately $10^{-11}/\sqrt{\tau \rm{(s)}}$ at 780 nm. A similar fit to the noise floor yields 1.49(8) kHz$/\sqrt{\tau \rm (s)}$. The Allan deviations of the locked lasers reach their minima at $\tau \simeq 10$ s, giving a frequency resolution of 1.9 kHz for follower 1 and 2.8 kHz for follower 2. At larger integration times ($\tau > 100$ s), the Allan deviations reaches the level of the noise floor and becomes independent of integration time (the so-called ``flicker floor''). At this level, the flicker frequency noise of the FVC at low frequencies limits further improvements in stability. Compared to the free-running case, the locking systems suppress laser frequency drift by more than $\times 600$ for $\tau > 1000$ s.

\begin{table}[!tb]
\centering
\begin{tabular}{cccc}
   \hline \hline
             & Capture     & FFI        & Dynamic    \\
   Reference & Range (MHz) & Resolution & Range (dB) \\
   \hline
   Stace et al.~\cite{Stace1998}       &    8 & $1.8 \times 10^{-13}$ & 50.5 \\
   McFerran et al.~\cite{McFerran2018} &   59 & $1.3 \times 10^{-13}$ & 62.3 \\
   This work (follower 1)              &  936 & $4.9 \times 10^{-12}$ & 57.0 \\
   This work (follower 2)              & 1424 & $7.3 \times 10^{-12}$ & 56.9 \\
   \hline \hline
\end{tabular}
\caption{Performance comparison of standalone locking systems based on a FVC IC. The FFI resolution is the minimum of the corresponding Allan deviation. The dynamic range is given by the capture range divided by the frequency resolution.}
\label{tab:Comparison}
\end{table}

Table \ref{tab:Comparison} shows a comparison of performance metrics for similar standalone locking systems. In our system, the capture range and FFI resolution are determined by the variable division ratio $D$. We find that the dynamic range (the ratio between these quantities) is approximately constant across a wide range of operating conditions. Compared to other work, we achieve a competitive dynamic range (within a factor of 4) for capture ranges up to 1.4 GHz.

%===========================================================================
\section{Cold-atom spectroscopy}
\label{sec:Spectroscopy}

\begin{figure*}[!tb]
  \centering
  \includegraphics[width=0.98\textwidth]{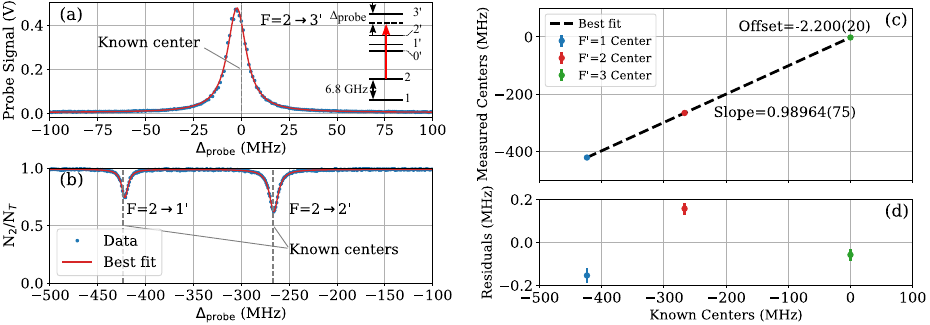}
  \caption{Spectroscopy with laser-cooled $^{87}$Rb. (a) Emission spectrum of the $F=2 \to F'=3$ cycling transition. Data are fit to a Lorentzian function. Inset shows the probe laser relative to the 5S$_{1/2}$ and 5P$_{3/2}$ energy levels. (b) Optical pumping spectrum of the $F=2 \to F'=1,2$ transitions. Data are fit to a sum of two Lorentzian functions. Vertical dashed lines indicate the known centers of each transition. (c) Comparison between the measured and known centers of each transition. A linear fit shows good agreement with the data. (d) Residuals of the linear fit in (c).}
  \label{fig:Spectroscopy}
\end{figure*}

To demonstrate the agility of our locking system, we measured several Doppler-free spectra with our cold $^{87}$Rb sample, as shown in \Fig \ref{fig:Spectroscopy}. We also use these spectra to verify the accuracy of the closed-loop frequency calibration. The following measurements were obtained from a sample of $\sim 10^8$ atoms loaded from a 2D MOT into a 3D MOT and cooled to $\sim 10~\mu$K using an optical molasses technique. The repetition rate of the experiment is approximately 2 Hz, limited primarily by the 300 ms loading of the 3D MOT. The magnetic field at the location of the atoms is probed using microwave spectroscopy. Our experiment is not magnetically shielded, so we typically reduce the field strength to $|B| < 50$ mG once per day using three mutually orthogonal pairs of Helmholtz coils. Before each spectroscopy measurement, atoms are prepared in the $F = 2$ ground state using an optical ``depumping'' pulse near the $F = 1 \to F'=2$ transition. All stages of this experiment utilized dynamic control of the follower 2 laser frequency enabled by our locking system \cite{Shalaby2025}.

Figure \ref{fig:Spectroscopy}(a) shows an emission spectrum of the $F = 2 \to F'=3$ transition in $^{87}$Rb. Here, we measure the atomic fluorescence induced by a short probe pulse (100 $\mu$s) as a function of the probe detuning $\Delta_{\rm probe}$ from the $F = 2 \to F'=3$ resonance. Each point on these spectra represents the light intensity measured by a photodiode during one repetition of the experiment. The probe, which is derived from follower 2, is swept from $\Delta_{\rm probe} = -100$ to 100 MHz in steps of 1 MHz using the control voltage. We convert these voltages to a detuning using the beat note calibration shown in \Fig \ref{fig:LockCalibration} and the known location of the primary laser. A least-squares fit to these data gave a line center of $-2.29(3)$ MHz.

Figure \ref{fig:Spectroscopy}(b) shows spectra of the $F = 2 \to F'=1,2$ transitions obtained using optical pumping spectroscopy. Here, we scanned the frequency of a weak probe pulse ($\sim 8$ $\mu$W, 100 $\mu$s) from $\Delta_{\rm probe} = -500$ MHz to $-100$ MHz. When the probe is near one of these resonances, atoms are efficiently pumped into the $F = 1$ ground state. We then measure the population remaining in $F = 2$ using a sequence of two ``detection'' pulses: the first is fixed at the $F = 2 \to 3$ resonance, and the second contains an additional frequency component (provided by an EOM sideband) near the $F = 1 \to 2$ resonance. The first detection pulse provides a fluorescence signal proportional to the number of atoms in $F = 2$ (denoted $N_2$). Similarly, the second pulse gives a signal proportional to the total number of atoms ($N_{\rm T}$). The atomic population in $F = 2$ is given by the ratio $N_2/N_{\rm T}$, which suppresses atom number fluctuations in the MOT and improves the signal-to-noise ratio of the spectra. Figure \ref{fig:Spectroscopy}(b) shows fits to these data, which yielded line centers of $-421.60(4)$ MHz and $-265.96(3)$ MHz for the $F = 2 \to 1$ and $F = 2 \to 2$ transitions, respectively. The relatively small uncertainties in these line centers ($\sim 30$ kHz) are a direct result of the frequency stability of our lock and the narrow Doppler width of the atomic cloud ($<100$ kHz FWHM).

To determine the accuracy of the control voltage mapping to optical frequency, we plot the measured line centers against their known values \cite{Steck2025}. An ideal calibration would give a linear relationship with unity slope and zero offset. Figure \ref{fig:Spectroscopy}(c) shows a linear fit to the data, which yields a slope of $0.98964(75)$ MHz/MHz and an offset of $-2.200(20)$ MHz. The relatively large offset is a result of the lock point of the primary laser. Latency introduced by a commercial lock-in amplifier causes the error signal to cross zero at a frequency shifted from the peak of the $F = 2 \to 3$ saturated absorption spectrum \cite{Shalaby2025}. The slope differs from unity by $\sim 1\%$, which is much larger than expected given the uncertainty in our closed-loop calibration ($\sim 0.01\%$) and the measured non-linearity ($< 0.1 \%$). This can be explained by the steady-state error of the lock on short timescales as follows.

\begin{figure}[!tb]
  \centering
  \includegraphics[width=0.48\textwidth]{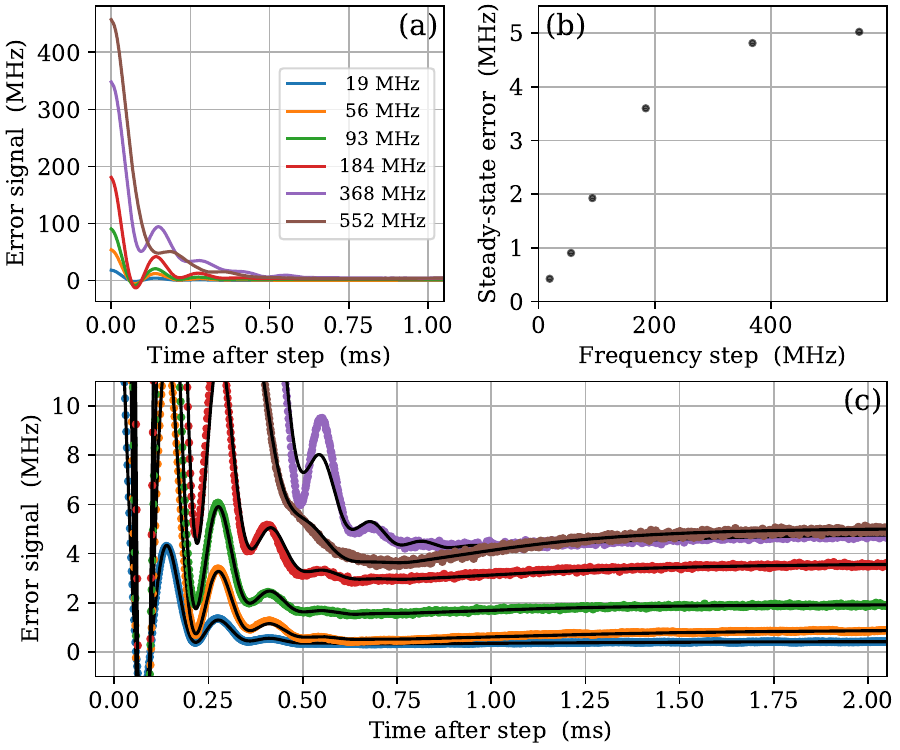}
  \caption{Lock agility measurements using the impulse response method. Here, the frequency of follower 1 is abruptly changed using a step in the control signal. (a) Error signal (converted to beat frequency units) as a function of time after the step. The legend indicates the magnitude of frequency steps. (b) Steady-state error as a function of the magnitude of frequency step. Values were determined from fits to the data shown in (c). Fit uncertainties are too small to be seen. (c) Zoom of impulse response data. Least-squares fits to each data set are shown as solid black lines and follow the functional form: $a_0 + a_1 e^{-\gamma_1 t} + a_2 (\gamma_2 t) e^{-\gamma_2 t} + a_3 (\gamma_3 t)^2 e^{-\gamma_3 t} + a_4 \cos(\omega_r t + \phi) e^{-\gamma_d t}$. The parameter $a_0$ is the steady-state error. The ring frequency is $\omega_r/2\pi = 7.14(1)$ kHz, the decay rate is $\gamma_d/2\pi = 1.97(1)$ kHz, and the $-3$~dB lock bandwidth is $f_{\rm bw} = (\omega_r^2 + \gamma_d^2)^{1/2}/2\pi = 7.40(1)$ kHz.}
  \label{fig:Agility}
\end{figure}

To scan the frequency of the probe pulse during this experiment, we step the control voltage from a fixed value (corresponding to the depump frequency) to the desired value for spectroscopy. The lock responds to this rapid change in the set point by undergoing damped oscillations, shown in \Fig \ref{fig:Agility}(a), as the PI controller works to stabilize the laser frequency. We wait 2 ms for the lock to stabilize before applying the probe pulse. However, because the fast integrator gain saturates at low frequencies, a small steady-state error persists. Importantly, the magnitude of this error increases with frequency step size [see \Figs \ref{fig:Agility}(b) and (c)], and its sign is independent of the direction of the frequency step. Thus, as the probe detuning is scanned, the atoms experience a larger detuning ($|\Delta_{\rm probe}|$) than expected, which effectively compresses the horizontal axis of the spectrum [\Figs \ref{fig:Spectroscopy}(a) and (b)] and causes the slope in \Fig \ref{fig:Spectroscopy}(c) to be less than one. The steady-state error reaches $\sim 1\%$ for frequency steps of 400 MHz, which agrees with the measured slope deviation from unity. This issue can be mitigated by using a slow integrator with larger gain at frequencies below 1 kHz. Presently, our slow integrator reduces the steady-state error below 1\% at long timescales ($> 67$ s), which was too slow for these spectroscopy experiments.

%===========================================================================
\section{Conclusion}
\label{sec:Conclusion}

We have presented a standalone, modular, low-noise frequency-offset locking system that leverages commercial electronic components to achieve agile, robust, and cost-effective control of narrow-linewidth lasers. Our architecture combines broadband frequency division, a highly linear FVC stage, and a PI controller to deliver GHz-scale capture ranges, ms-scale response times, and kHz-level frequency stability without the need for active photonic components, an ultra-stable local oscillator, or digitally generated RF sources. Closed-loop characterization demonstrates excellent linearity and strong suppression of low-frequency noise, reaching an FFI resolution of $4.9 \times 10^{-12}$ (1.9 kHz at 780 nm). High-resolution spectroscopy of laser-cooled $^{87}$Rb atoms further validates the tunability and precision of the locking system. These results illustrate that FVC-based locks remain a powerful and scalable solution for atomic physics experiments. Their large capture range makes them ideal for the first stage of an OPLL. Compared to similar standalone systems \cite{Stace1998, McFerran2018}, our lock achieves competitive dynamic range for capture ranges up to 1.4 GHz. Our design also allows for dynamic adjustment of the division ratio, which can offer improved resolution at the expense of smaller capture range. The present implementation is limited by the bandwidth of the feedback loop, the stability of the control voltage, and the temperature sensitivity of circuit components.

Although the FVC used in this work (AD650) is nearing the end of its lifecycle, several alternatives exist, such as the Texas Instruments VFC320 (1 MHz full-scale range, 0.1\% nonlinearity at 1 MHz) and the VFC110 (4 MHz full-scale range, 0.02\% nonlinearity at 2 MHz). Improved accuracy and stability was demonstrated by Reynolds et al.\cite{Reynolds2019} using a clock-synchronized FVC, such as the Analog Devices AD652S. Other enhancements may include higher-bandwidth FVC designs \cite{Bui2005}, or a fully digital implementation. While these options would increase the design complexity and resources required (e.g., a stable clock source), they promise faster dynamic control and extend locking capabilities to larger beat-note frequencies.

%===========================================================================

\begin{acknowledgments}

The authors gratefully acknowledge funding from the Natural Science and Engineering Research Council (NSERC, grant no.~RGPIN-2021-02629), the New Brunswick Innovation Foundation (NBIF, grant no.~RAI-2022-054), the Canadian Foundation for Innovation (CFI, grant no.~41597), the Innovation for Defence Excellence and Security (IDEaS, grant no.~MN4-032), the Harrison McCain Foundation, and the University of New Brunswick (UNB). We thank Brian Titus of UNB's Physics Department Machine Shop for his assistance constructing components of our experiment. We also thank Prof.~P.T. Jayachandran of UNB's Radio Space Physics Laboratory for the generous loan of a high-bandwidth spectrum analyzer early in this study.

\end{acknowledgments}

%===========================================================================

\section*{Author Declarations}

\subsection*{Conflicts of Interest}

The authors have no conflicts of interest to disclose.

\subsection*{Author Contributions}

\noindent \textbf{K. Shalaby}: Conceptualization (supporting); Formal analysis (lead); Investigation (lead); Methodology (lead); Software (lead); Writing – original draft (lead); Writing – review \& editing (supporting). \textbf{T. Hunt}: Formal analysis (supporting); Investigation (supporting); Software (supporting); Writing – review \& editing (supporting). \textbf{S. Moir}: Investigation (supporting); Writing – review \& editing (supporting). \textbf{P. Trottier}: Conceptualization (supporting); Investigation (supporting); Methodology (supporting); Resources (supporting); Supervision (supporting); Writing – review \& editing (supporting). \textbf{T. Reuschel}: Conceptualization (supporting); Investigation (supporting); Methodology (supporting); Supervision (supporting); Writing – review \& editing (supporting). \textbf{B. Barrett}: Conceptualization (lead); Formal analysis (supporting); Funding acquisition (lead); Methodology (supporting); Project administration (lead); Resources (lead); Software (supporting); Supervision (lead); Writing – original draft (supporting); Writing – review \& editing (lead).

%===========================================================================

\section*{Data Availability}

The data that support the findings of this study are available from the corresponding author upon reasonable request. Our circuit designs for the frequency-offset locking system are openly available from the following repository: 10.5281/zenodo.18876772.

%===========================================================================

\bibliography{Bibliography_FrequencyOffsetLock}

%%%%%%%%%%%%%%%%%%%%%%%%%%%%%%%%%%%%%%%%%%%%%%%%%%%%%%%%%%%%%%%%%%%%%%%%%%%%
\end{document}